\documentclass[sigconf]{acmart}

\setcopyright{none} 

\acmConference{}{}{}
\renewcommand\footnotetextcopyrightpermission[1]{}

\acmYear{2022}
\usepackage{listings}
\usepackage{mdframed}
\mdfdefinestyle{listingframe}{%
}

\usepackage{xspace}
\usepackage{url}
\usepackage{hyperref}
\usepackage{makecell}
\usepackage{amsmath}
\usepackage{tikz}
\usepackage{graphicx}
\usepackage{enumitem}

\usepackage{caption}
\usepackage{subcaption}

\usepackage[section]{placeins}

\usepackage{color}
\definecolor{halfred}{HTML}{800000}
\definecolor{halfgreen}{HTML}{008000}
\definecolor{halfblue}{HTML}{008080}
\definecolor{codeindent}{HTML}{cccccc}

\def\mw/{\emph{Microwalk}\xspace}
\def\mwci/{\emph{Microwalk-CI}\xspace}
\def\jalangitwo/{Jalangi2\xspace}
\def\DATA/{\emph{DATA}\xspace}
\def\javascript/{JavaScript\xspace}
\def\contanttime/{constant-time\xspace}
\def\sidechannel/{side-channel\xspace}
\def\clanguage/{C\xspace}
\def\cpplanguage/{C++\xspace}
\def\controlflow/{control flow\xspace}
\def\dataflow/{data flow\xspace}
\def\npm/{NPM\xspace}
\def\nodejs/{Node.js\xspace}

\usepackage{todonotes}

\newcommand{\bheading}[1]{{\vspace{4pt}\noindent{\textbf{#1}}}}

\usepackage[nolist]{acronym}
\begin{acronym}
  \acro{IoT}{Internet of Things}
  \acro{LLC}{Last Level Cache}
  \acro{CI}{Continuous Integration}
  \acro{JIT}{just-in-time}
\end{acronym}  

\AtBeginDocument{%
  \providecommand\BibTeX{{%
    \normalfont B\kern-0.5em{\scshape i\kern-0.25em b}\kern-0.8em\TeX}}}

\begin{document}

\title{Microwalk-CI: Practical Side-Channel Analysis for JavaScript Applications}

\author{Jan Wichelmann}
\affiliation{%
  \institution{University of Lübeck}
  \city{Lübeck}
  \country{Germany}}
\email{j.wichelmann@uni-luebeck.de}

\author{Florian Sieck}
\affiliation{%
  \institution{University of Lübeck}
  \city{Lübeck}
  \country{Germany}}
\email{florian.sieck@uni-luebeck.de}

\author{Anna Pätschke}
\affiliation{%
  \institution{University of Lübeck}
  \city{Lübeck}
  \country{Germany}}
\email{a.paetschke@uni-luebeck.de}

\author{Thomas Eisenbarth}
\affiliation{%
  \institution{University of Lübeck}
  \city{Lübeck}
  \country{Germany}}
\email{thomas.eisenbarth@uni-luebeck.de}

\begin{abstract}
Secret-dependent timing behavior in cryptographic implementations has resulted in exploitable vulnerabilities, undermining their security. Over the years, numerous tools to automatically detect timing leakage or even to prove their absence have been proposed. 
However, a recent study at IEEE S\&P 2022  showed that, while many developers are aware of one or more analysis tools, they have major difficulties integrating these into their workflow, as existing tools are tedious to use and mapping discovered leakages to their originating code segments requires expert knowledge. In addition, existing tools focus on compiled languages like C, or analyze binaries, while the industry and open-source community moved to interpreted languages, most notably JavaScript.

In this work, we introduce Microwalk-CI, a novel side-channel analysis framework for easy integration into a JavaScript development workflow. First, we extend existing dynamic approaches with a new analysis algorithm, that allows efficient localization and quantification of leakages, making it suitable for use in practical development. We then present a technique for generating execution traces from JavaScript applications, which can be further analyzed with our and other algorithms originally designed for binary analysis. Finally, we discuss how  Microwalk-CI can be integrated into a continuous integration (CI) pipeline for efficient and ongoing monitoring. We evaluate our analysis framework by conducting a thorough evaluation of several popular JavaScript cryptographic libraries, and uncover a number of critical leakages.
\end{abstract}

\keywords{}

\maketitle

\section{Introduction}

Collection of sensitive data is common in today's cloud and \ac{IoT} environments, and affects everyone. Protecting this private and sensitive data is of utmost importance, therefore requiring secure cryptography routines and secrets. However, especially the cloud allows attackers to observe the execution of victim code using side-channels in co-located environments~\cite{DBLP:conf/ches/InciGIES16}.
These attacks range from \ac{LLC}~\cite{DBLP:conf/sp/LiuYGHL15} and de-duplication attacks~\cite{DBLP:conf/sac/LindemannF18} to the observation of memory access patterns~\cite{DBLP:conf/sp/XuCP15} or main memory access contention~\cite{DBLP:conf/uss/PesslGMSM16}. The spatial resolution depends on the granularity of the attacked buffer and the temporal resolution on the capabilities of the attacker, meaning either the ability to achieve a sufficiently high measurement frequency~\cite{DBLP:conf/uss/YaromF14} or to interrupt and pause the victim code~\cite{vanbulck2017SGXStep,DBLP:conf/sp/XuCP15}.

To avoid side-channel vulnerabilities, programmers should write \contanttime/ code, i.e., software which does not contain input or secret-dependent \controlflow/ or memory accesses.
Depending on the problem at hand, this can be achieved by different means: For example, a secret-dependent data access may be replaced by accessing every element of the target array and then choosing the correct one with a mask. Conditionals can be adjusted by always executing both branches and then selecting the result.

However, for complex projects like large cryptographic libraries, finding such vulnerabilities is a difficult and time-intensive task. 
Thus, the research community has developed a number of analysis strategies~\cite{langley2010ctgrind,DBLP:conf/uss/DoychevFKMR13,DBLP:conf/date/ReparazBV17,DBLP:conf/uss/AlmeidaBBDE16,DBLP:conf/uss/WangWLZW17,weiser2018data,wichelmann2018microwalk,DBLP:conf/sp/BrotzmanLZTK19,DBLP:conf/sp/DanielBR20}, that aim at automating the detection of side-channel leakages in a given code base. 
However, a recent study~\cite{jancar2022mitigate} that conducted a survey between crypto library developers found that while most developers were aware of and welcome those tools, they had major difficulties using them due to bad usability, lack of availability and maintenance or high resource consumption. 
The authors worked out a number of recommendations for creators of analysis tools: The tools should be well-documented and easily usable, such that adoption requires low effort from the developer. 
Another focus is on compatibility: The analysis shouldn't require use of special languages or language subsets. 
Finally, the tools should aid \emph{efficient} development, i.e., quickly yield results with less focus on completeness, making them suitable for inclusion in a continuous integration (CI) workflow.

In this work, we study how these challenges can be addressed, and adapt the existing \mw/~\cite{wichelmann2018microwalk} framework to fit the given objectives. 
\mw/ was originally designed for finding leakages in binary software, for which it generates a number of execution traces for a set of random inputs and then compares them.
The dynamic analysis approach of \mw/ is quite fast, as it does only run the target program several times, and then compares the resulting execution traces with a simple linear algorithm. 
However, due to the simplistic leakage quantification, the resulting analysis reports contain a lot of potential vulnerabilities, with little or even misleading information about their cause. 
This makes it difficult to assess their severity and address them efficiently, especially for complex libraries. Finally, the initial setup can be time-consuming, as the different components need to be compiled from source.

We mitigate these issues by designing \mwci/, which features a new leakage analysis algorithm that combines the performance benefits of dynamic analysis with an accurate leakage localization and quantification, easing the assessment and investigation of the reported leakages. In addition, we add support for running \mwci/ in an automated environment like a CI pipeline, and create a Docker image that contains \mwci/ and its dependencies for easy use. Finally, we create simple templates
that allow quick adoption of \mwci/'s analysis capabilities into a cryptographic library's CI workflow with little effort by the developer.

During the research on leakage detection tools, as part of their evaluation, many vulnerabilities in popular cryptographic libraries have been uncovered and fixed.
However, the developer community is moving away from compiled languages like C or C++ and instead embraces interpreted scripting languages like \javascript/ or Python. 
In fact, the 2021 Stack Overflow developer survey and the January 2021 Redmonk programming language ranking found that those two languages are the most popular, both for private and professional contexts~\cite{stackoverflow-developer-survey, redmonk-language-ranking}. 
\javascript/ was originally designed as a client-side language for web browsers, but, with the arrival of \nodejs/~\cite{nodejs-tool}, it has seen growing adoption for server-side software as well. 
Consequently, the community has come up with a number of cryptographic libraries written in pure \javascript/. 
However, due to the lack of appropriate tooling and attention of the research community, these libraries have never been vetted for their robustness against side-channel attacks, which is worrying given the fact that the servers using them may be hosted in IaaS cloud environments.

To address this, \mwci/ offers a novel method for applying \mw/'s original binary analysis algorithms to \javascript/ libraries by using the \jalangitwo/~\cite{jalangi2github,jalangi2tutorial} source code instrumentation library to generate compatible traces.
The new tracing backend comes with a simple code template and supports full automation, such that the analysis can be easily added to the CI workflows of respective libraries. We evaluate several popular \javascript/ cryptographic libraries, uncovering a number of high-severity leakages.

By supporting the analysis of \javascript/, we strive to improve the security of software and rise awareness for the importance of constant-time cryptographic code in the community of web and cloud developers. The underlying concepts of our source-based trace generator can be used for building analysis support for other programming languages as well, making side-channel leakage analysis available for all common platforms and at a low barrier.

\subsection{Our Contribution}
In summary, we make the following contributions:

\begin{itemize}
    \item We introduce a novel call tree-based analysis method, which allows efficient and accurate localization and quantification of leakages.
    \item We show the first dynamic leakage analysis tool for \javascript/ libraries.
    \item We propose a new approach for integrating a fully automated timing leakage analysis into the crypto library development workflow, which requires low effort from the developer and immediately reports newly introduced vulnerabilities.
    \item We evaluate the new analysis framework with several widely-used \javascript/ libraries and uncover a significant number of previously unaddressed leakages.
\end{itemize}

The source code of \mwci/ is available at \url{https://github.com/microwalk-project/Microwalk}.

\subsection{Disclosure}
We contacted the authors of the affected libraries, informed them about our findings and offered to aid in fixing the vulnerabilities.

The author of \texttt{elliptic} acknowledged the discovered vulnerabilities, but noted that the package is no longer maintained and that fixing the vulnerabilities would require major changes, as side-channel resistance wasn't part of the underlying design considerations. There was no response from the other library authors.

\section{Background}
\label{sec:background}

\subsection{Microarchitectural Timing Attacks}
Implementations of cryptographic algorithms are often run on hardware resources that are shared between different processes.
If the code exhibits secret-dependent behavior, malicious processes can use the resulting information leakage to extract secrets like private keys through side-channel analysis.
Cache attacks are a prominent example for exploiting resource contention with a victim process: By measuring the time it takes for repeatedly clearing and accessing a specific cache entry, the attacker can see whether the victim accessed a similar cache entry in the meantime~\cite{bernstein2005cache,DBLP:conf/ctrsa/OsvikST06,DBLP:conf/ches/AciicmezBG10,DBLP:conf/uss/YaromF14,DBLP:conf/uss/ShustermanAOGOY21}.
Other attack vectors include the translation lookaside buffer (TLB)~\cite{DBLP:conf/uss/GrasRBG18} and the branch prediction unit~\cite{DBLP:conf/ctrsa/AciicmezKS07}.

The most widely used software countermeasure against these attacks is writing constant-time code that does not contain secret-dependent memory accesses or branches, and that uses instructions that do not come along with operand-dependent runtime~\cite{DBLP:journals/iacr/BrickellGNS06}.
There exists a variety of tools~\cite{langley2010ctgrind,DBLP:conf/uss/DoychevFKMR13,DBLP:conf/date/ReparazBV17,DBLP:conf/uss/AlmeidaBBDE16,DBLP:conf/uss/WangWLZW17,weiser2018data,wichelmann2018microwalk,DBLP:conf/sp/BrotzmanLZTK19,DBLP:conf/sp/DanielBR20}, that feature different analysis approaches. Some of these tools are open-source, with varying performance and usability~\cite{jancar2022mitigate}.

\subsection{Microwalk}
\mw/~\cite{wichelmann2018microwalk} is a framework for checking the constant-time properties of software binaries in an automated fashion. It follows a dynamic analysis approach, i.e., it executes the target program with a number of random inputs and collects execution traces, which contain branch targets, memory allocations and memory accesses. This is done through a three-stage pipeline, where traces are generated, preprocessed, and analyzed. Each stage has various \emph{modules}, which are chosen by the user depending on their application. Furthermore, \mw/ has a plugin architecture, that allows easy extension by loading custom modules.

Currently, \mw/ only has one trace generation module, which is based on Intel Pin~\cite{intelpin-user-guide} and produces traces for binary software. Correspondingly, there is a preprocessor module that converts the raw traces generated by Pin into \mw/'s own format. Finally, these preprocessed traces can be fed into a number of analysis modules, e.g., for computing the mutual information between memory access patterns and inputs, or for dumping the preprocessed traces in a human-readable format.

\subsection{Mutual Information and Guessing Entropy}
Mutual information (MI) quantifies the interdependence of two random variables, i.e., it models how much information an attacker can learn about one variable on average by observing the other one~\cite{guiasu1977information}. It has been widely used for quantifying side-channel leakages~\cite{DBLP:conf/acsac/ZhangL14,DBLP:journals/tc/BayrakRNBSI15,DBLP:journals/corr/abs-1709-01552,wichelmann2018microwalk}.

The mutual information of the random variables $X\colon K\rightarrow\mathcal{X}$ and $Y\colon L\rightarrow\mathcal{Y}$ is defined as
\begin{align*}
    I(X,Y)&=\sum_{\substack{x \in \mathcal{X}\\y \in \mathcal{Y}}}\Pr[X=x,Y=y]\cdot\log_2\left(\frac{\Pr[X=x,Y=y]}{\Pr[X=x]\cdot \Pr[Y=y]}\right).
\end{align*}
The information is measured in bits. In our setting, the random variable $X$ represents a secret and $Y$ the information that can be gathered by observing the system state through a side-channel.

The guessing entropy (GE) of a random variable $X\colon K\rightarrow\mathcal{X}$ quantifies the average number of guesses that have to be made in order to guess the value of $X$ correctly~\cite{DBLP:conf/ccs/KopfB07}. If $\mathcal{X}$ is indexed such that $\Pr[X=x_i]\geq\Pr[X=x_j]$ for $x_i,x_j\in\mathcal{X}$ and $i\leq j$, the guessing entropy is defined as
\begin{align*}
    G(X)&=\sum_{1\leq i\leq|\mathcal{X}|}i\cdot\Pr[X=x_i].
\end{align*}
The \emph{conditional guessing entropy} (conditional GE) $G(X\,|\,Y)$ for random variables $X$ and $Y$ is defined as
\begin{align*}
    G(X\,|\,Y)&=\sum_{y \in \mathcal{Y}} \Pr[Y=y] \cdot G(X\,|\,Y=y).
\end{align*}
$G(X\,|\,Y)$ measures the expected number of guesses that are needed to determine the value of $X$ for a known value of $Y$.

A variant of the conditional GE, the \emph{minimal conditional guessing entropy} (minimal GE), determines the lower bound of expected guesses.
It is defined as
\begin{align*}
    \hat{G}(X\,|\,Y)&=\min_{y \in \mathcal{Y}} G(X\,|\,Y=y),
\end{align*}
i.e., it outputs the minimal number of guesses that are needed to find out one of the possible values of $X$.

\subsection{\javascript/ Instrumentation}

\javascript/ code can be instrumented in different ways, each coming with their own benefits and drawbacks. 

FoxHound~\cite{sapprojectfoxhound} modifies Firefox's \javascript/ engine. %
While this allows many optimizations, it comes with the downside of being constrained to one specific \javascript/ engine and requiring constant maintenance to keep up with the upstream project. OpenTelemetry~\cite{opentelemetryjs} and Google's tracing framework~\cite{googletracingframework} create program traces to monitor and profile software, but require the developer to insert instrumentation calls into their source code manually.
While being very specific and thus only introducing the necessary overhead, they are not generally applicable
without a lot of manual effort.

Lastly, the \javascript/ code can be dynamically instrumented in a source-to-source fashion. \jalangitwo/~\cite{DBLP:conf/sigsoft/SenKBG13,jalangi2github,jalangi2tutorial} wraps the loading process of \javascript/ files
and injects instrumentation code into the source code.
The user of the instrumentation framework can write and register custom callback routines, which are supplied with the current execution state.
This approach comes with a certain overhead, but it is flexible and works with arbitrary \javascript/ code without manual adjustments.

\section{A Fast Leakage Analysis Algorithm}
\label{sec:leakage-analysis}
We propose a new leakage analysis algorithm that is optimized for quickly delivering detailed leakage information, aiding developers in efficiently locating and fixing issues. Before we dive into the algorithm, we define the leakage model and discuss the objectives a thorough leakage analysis must meet. Then, we describe how the traces are processed to build a call tree, which in a final step is broken down to compute leakage metrics for specific instructions.

\subsection{Leakage Model}
To ensure that we detect all leakages which may be exploited by current and future attack methods, we choose a strong leakage model: An  attacker tries to extract secret inputs from an implementation through a side-channel attack, which allows them to get a trace of all executed instructions and all accessed memory addresses. They also have access to all public inputs and outputs.

Under certain conditions, a hypervisor/OS-level adversary can single step instructions~\cite{DBLP:journals/tches/AldayaB20,DBLP:conf/ccs/SieckBW021,DBLP:conf/uss/MoghimiBHPS20}, or have below cache-line resolution~\cite{DBLP:conf/ches/YaromGH16,DBLP:journals/ijpp/MoghimiWES19}. However, for more relaxed adversarial scenarios like cross-VM attacks, granularities of 32 or 64 bytes and hundreds of instructions may be more appropriate. Adjusting the processing of the leakage accordingly allows an analysis under such a leakage model as well, but, we believe that the most conservative approach should be applied, i.e., assuming a maximum resolution attacker.
Attacks exploiting speculative execution are considered off-scope, as we focus on leakages caused by actual secret-dependent control flow or memory accesses, i.e., code paths that are reached architecturally.

This leakage model and the following analysis approach are consistent with the models used by \mw/~\cite{wichelmann2018microwalk} and \DATA/~\cite{weiser2018data}.

\subsubsection{Analysis approach}
The leakage model can be turned into a dynamic analysis approach by making the following observation: Since the attacker tries to infer a secret solely by looking at an execution trace and public inputs/outputs, they can only succeed if the trace depends on the secret. I.e., if changing the secret does never influence the observed trace, the implementation does not leak the secret and is \emph{constant-time}.

We model this by giving the attacker a number of secret inputs and corresponding execution traces, and asking them to map the inputs to the respective traces. If they perform better than guessing, we consider the implementation as leaking. If all traces are identical, the implementation is considered constant-time.

\subsection{Objectives}
For an efficient and useful dynamic leakage analysis, we identified three major objectives: Accurate \emph{localization} of leakages, a \emph{quantification} of leakage severity, and \emph{performance}.

\paragraph{Localization} While varying address traces for a memory read instruction are a clear sign that there \emph{is} leakage, which can be extracted by monitoring that particular instruction~\cite{wichelmann2018microwalk}, they do not indicate where the leakage is actually \emph{caused}. E.g., a non-constant-time function may be called two times, once with a secret-dependent parameter, and once a varying number of times in a loop, but with a constant parameter (Figure~\ref{fig:varying-loop-iterations}). A correct analysis should distinguish the two invocations of \texttt{lookup} and mark the table access in line 12 as leaking for the first invocation (line 2); for the second invocation (line 6), the secret-dependent branch in line 5 should be reported, as the table access in line 12 itself does not add any leakage.

\begin{figure}[t]
\begin{mdframed}[style=listingframe]
\begin{lstlisting}[language=C,numbers=left,xleftmargin=2em]
int func(int secret) {                   // func+0
    lookup(secret);                      // func+1

    int result = 0;
    for(int i = 0; i < secret; ++i) {    // func+4
        result += lookup(1);             // func+5
    }                                    // func+6
    return result;                       // func+7
}
int table[] { ... };
int lookup(int index) {                  // lookup+0
    return table[index];                 // lookup+1
}
\end{lstlisting}
\end{mdframed}
\caption{A sample program illustrating different kinds of leakages: The \texttt{lookup} function is not constant-time, since it does an input-based array lookup, so the memory access to \texttt{table[index]} would be marked as leaking if \texttt{index} is secret. Another cause of leakage is in \texttt{func}, which calls \texttt{lookup} a varying number of times depending on a secret value.}  \label{fig:varying-loop-iterations}
\end{figure}

\paragraph{Quantification} In addition to an accurate localization, there is a need for a rough quantification of the severity of leakages. For example, a chain of nested \texttt{if} statements may only leak a few bits of the secret each, but the leakage aggregates up to a point which allows an attacker to easily distinguish different secrets just by looking at the resulting sequence of branch instructions. At the same time, a lone \texttt{if} statement which merely handles a special case during key file parsing (e.g., whether a parameter has some additional byte) does not necessarily pose an urgent problem. The analysis should assign each leakage with a score allowing the developer to prioritize between findings.

\paragraph{Performance} Finally, for integrating the leakage analysis into a development workflow, performance is important: When checking whether a proposed change impacts security, or whether a given patch fixes a previously discovered leakage, the developer should not need to wait several ten minutes or hours until analysis results are available. The analysis should be efficient enough to run it both on a standard developer machine and in a hosted CI environment.

\subsection{Algorithm Idea}
In order to find leakages, we need to compare the generated traces, and find sections where they diverge and, later, merge again. However, due to the performance requirements and the immense size of traces, especially for asymmetric primitives, we cannot afford running a traditional diff or trace alignment algorithm, which usually have quadratic complexity. At the same time, we do not want to lose information, as we want to accurately pinpoint the detected leakages. Thus, we opt for a data structure that preserves all necessary information in an efficient way, and which allows to conduct a thorough leakage analysis which can discover and quantify trace divergences in linear time. 

For that, we merge the traces into a call tree, where each function call and a few other trace entries form the nodes, and where subsequent function calls \emph{and trace divergences} generate branches. Each node holds the IDs of the traces which reach that node. The tree can be built on-the-fly while the traces are processed, so it can be integrated into a leakage analysis pipeline like the one offered by \mw/. After the traces have been processed, a final step traverses the tree and evaluates for each instruction in each call stack, whether it caused a divergence and how severe that divergence is.
In the following sections, we elaborate on the respective steps.

\subsection{Step 1: Building the Call Tree}
We merge the traces into a tree in a greedy way, i.e., we simultaneously iterate over a trace and the current tree entries, and add the trace entries to the tree. In order to save memory and get a readable representation of the traces with little tree depth, we can exploit the fact that traces of constant-time implementations tend to have long shared sequences without any differences, and thus use a radix trie instead of a plain tree, such that each node holds an as long as possible sequence of consecutive trace entries. %

The resulting tree for the example code in Figure~\ref{fig:varying-loop-iterations} is illustrated in Figure~\ref{fig:call-tree-dump} in Appendix~\ref{sec:app:call-tree-dump}.

\subsubsection{Types of trace entries}
In order to address the leakage model, the execution traces used by \mw/ contain information about branches, memory allocations and memory accesses.

\bheading{Branches}
cover call, return and jump instructions. A \emph{branch} trace entry has a source address, a target address and a \emph{taken} bit that denotes whether the branch was taken or skipped (e.g., due to a failed comparison). The source and target addresses consist each of an image ID (i.e., the binary which contains the corresponding instruction) and an offset.

\bheading{Memory allocations}
are used to keep track of memory blocks on the heap and stack. Each time the analyzed program calls \emph{malloc} or a similar function, a new allocation block is registered with a unique ID and the allocation block size.

\bheading{Memory accesses}
contain the image ID and offset of the corresponding instruction and the allocation block ID and offset of the accessed address. This relative addressing allows to compare traces even when they each operate on their own allocated memory regions, which have different absolute addresses.

\subsubsection{Tree layout}
As mentioned above, we chose a radix trie-like representation of the merged trace entries, as this reduces tree depth, speeds up analysis and enhances readability of tree dumps.

\begin{figure}[t]
\includegraphics[width=\columnwidth]{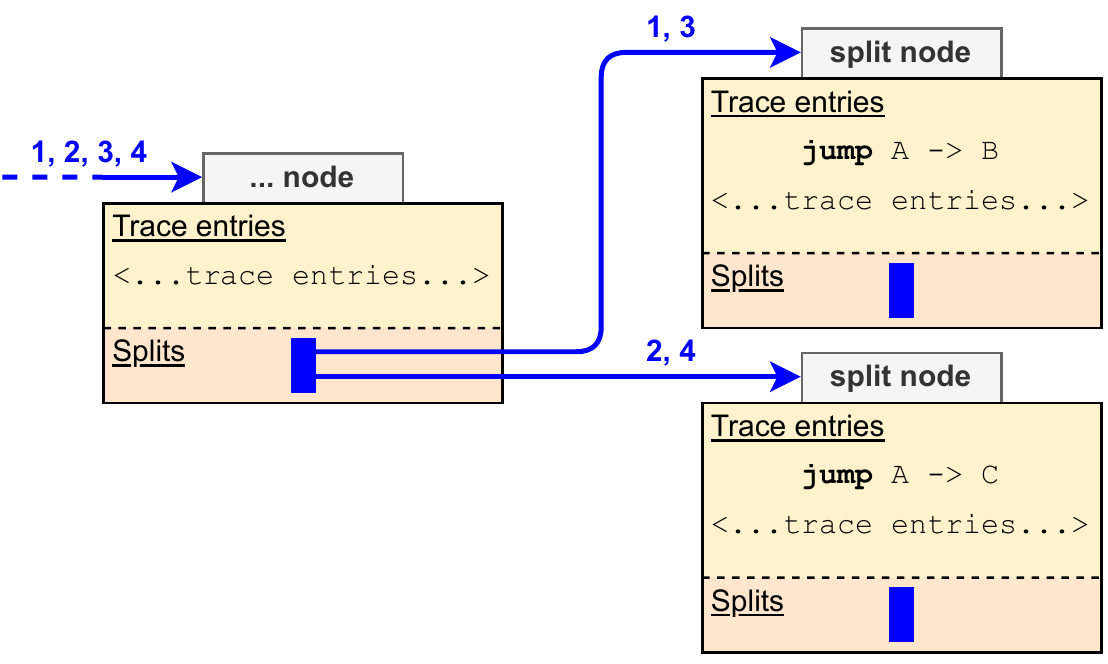}
\caption{A generic trace divergence with two split nodes. While traces 1 to 4 share the entries in the left node, they differ at the jump statement at location \texttt{A}: Traces 1 and 3 jump to location \texttt{B}, while traces 2 and 4 jump to \texttt{C}. Here, each case gets its own split node, and processing of trace entries is resumed there.}
\label{fig:split-node}
\end{figure}

\begin{figure}[t]
\includegraphics[width=\columnwidth]{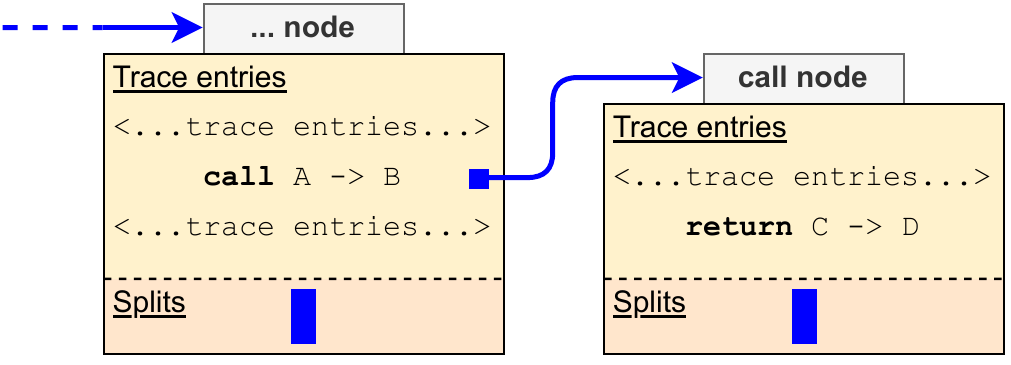}
\caption{A generic function call with a call node. When a function call entry is encountered, a new call node is created, that subsequently receives the trace entries for the given function. Once the function ends (return statement), the trace entry list of the prior call node is continued. Note that the return statement may also end up in the split node tree of the call node, if there are trace divergences within the function.}
\label{fig:call-node}
\end{figure}

A tree node consists of two parts: The consecutive trace entries which are present for all traces hitting this node, and a list of (edges to) \emph{split nodes} (Figure~\ref{fig:split-node}), which represent divergences between the different traces. The trace entry list may contain \emph{call nodes} (Figure~\ref{fig:call-node}) which open their own sub tree, but always return back into the current node and may be followed by other trace entries.

Edges start from within the trace entry list (for calls) or from the split node list. If an edge leads to a split node, it is annotated with the trace IDs taking this specific edge.

\subsubsection{Inserting trace entries into the tree}
The handling of equal and conflicting trace entries depends on the respective type. Split nodes are only created when a function call or a jump targets a different instruction than the already existing trace entry, as the resulting sub tree may be fairly different. Other differences like varying memory access offsets are only recorded in the respective trace entry, as they don't affect \controlflow/ and the current entry is thus likely followed by other, non-conflicting entries.

A function call is handled by creating a new tree node at the current position in the list of consecutive trace entries of the current tree node. Afterwards, the current node is pushed onto a stack and the new call node is set as the current node, such that subsequent trace entries are stored in the new node. When encountering a return statement, the last node is popped from the stack, and insertion of trace entries is resumed after the earlier created call node. If the target address of the current call entry does not match the target address recorded in an existing call node, a split is triggered.

If a conflict between an existing and a new trace entry is detected, the algorithm generates two new split nodes: One node receives the original conflicting trace entry, the remaining consecutive trace entries and the split node list of the current node; the other node is initialized with the new conflicting trace entry and an empty split node list. The branches to both nodes are annotated with the corresponding trace IDs. The current node is then set to the new split node, such that the new trace entries end up in the new node. The call node stack is \emph{not} updated, i.e., the next return statement ends the divergence and restores the state before the last call node. This way, we can recover from a trace divergence and discover additional leakages in other function calls.

Cases where there are more than two possible targets for an instruction (e.g., an indirect jump) are handled appropriately, by generating further split nodes at the same level.

\subsection{Step 2: Leakage Analysis}
After trace processing has concluded, we have a call tree that encodes the similarities and differences of all traces. We now perform a final step that collects this information and computes leakage measures, such that we can assign leakage information to each instruction, meeting our localization and quantification objectives.

\begin{figure}[t]
    \centering
    \begin{subfigure}[b]{0.4\textwidth}
        \centering
        \begin{mdframed}[style=listingframe]
        \begin{lstlisting}[language=C]
void f1(int secret) { f2(secret); }       // f1+0
void f2(int secret) { f3(secret); }       // f2+0
void f3(int secret) {                     // f3+0
    int tmp = 0;
    for(int i = 0; i < 2; ++i) {          // f3+2
        if(secret & (1 << i)) {           // f3+3
            ++tmp;
        }
    }                                     // f3+6
}
        \end{lstlisting}
        \end{mdframed}
        \caption{Program}
        \label{fig:trace-id-tree:code}
    \end{subfigure}
    
    \begin{subfigure}[b]{0.15\textwidth}
        \centering
        \begin{mdframed}[style=listingframe]
        \begin{lstlisting}[language=C]
(*@\textbf{Call stack:}@*)
  main+X -> f1+0
  f1+0 -> f2+0
  f2+0 -> f3+0

(*@\textbf{Instructions:}@*)
  jump at f3+2
  jump at f3+3
  jump at f3+6
        \end{lstlisting}
        \end{mdframed}
        \caption{Call stack and\\instruction info}
        \label{fig:trace-id-tree:call-stack}
    \end{subfigure}
    \hfill
    \begin{subfigure}[b]{0.3\textwidth}
        \centering
        \includegraphics[width=0.9\textwidth]{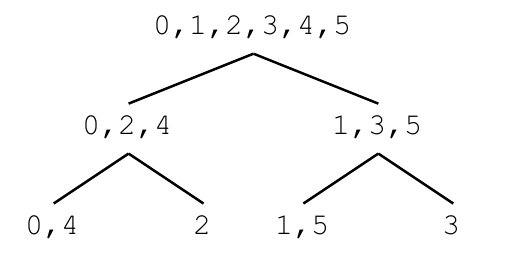}
        \caption{Trace ID tree for jump at \texttt{f3+3}}
        \label{fig:trace-id-tree:tree}
    \end{subfigure}
    \caption{Example for call stack and trace ID generation. The program in (a) counts the number of \texttt{1}s in the two least-significant bits of a secret variable by repeatedly executing a secret-dependent \texttt{if} statement. When calling \texttt{f1} from \texttt{main} with secret values from 0 (trace ID 0) to 5 (trace ID 5), we get the call stack as shown in (b), with three detected jump instructions. The secret-dependent \texttt{jump} at \texttt{f3+3} leads to divergence of traces, as is visible in the resulting trace ID tree in (c). Traces sharing a tree node at tree level $h\geq 0$ are identical for at least $h$ consecutive executions of the instruction.}
    \label{fig:trace-id-tree}
\end{figure}

\subsubsection{Building call stacks with trace ID trees per instruction}
\label{sec:leakage-analysis:leakage-analysis:building-trace-id-trees}
First, we consolidate the call tree into a number of call stacks, and store the trace split information for each instruction in the corresponding call stack. The split information consists of \emph{trace ID trees}, which encode how multiple executions of the given instruction for a certain function invocation led to trace divergence. This greatly simplifies the computation of leakage measures for individual instructions, and allows to display expressive information about the leakage behavior of a given instruction to the developer.
If a function is called multiple times (i.e., the same call stack occurs repeatedly), additional trace ID trees are created (no merging).

When a split is encountered, new child nodes for each edge of the split are added to the trace ID tree for the responsible instruction.
Figure~\ref{fig:trace-id-tree} illustrates the resulting trace ID tree for a simple program counting bits in a secret variable: When the jump instruction in question is encountered first, all traces are identical (tree level 0). At that point, execution diverges for traces with an even versus an odd secret. After the second iteration, traces are again split depending on the second bit of the secret. In the end, there are four different possible traces for the given function call.

\subsubsection{Computing leakage measures}
\label{sec:leakage_measures}
After recording the divergence behavior of instructions per call stack, we can compute various measures to quantify the corresponding leakage. We feature three efficiently computable metrics that give the developer an indication of the severity of each detected leakage: Mutual information, conditional guessing entropy and minimal conditional guessing entropy.
If the function containing the analyzed instruction is invoked multiple times for the same call stack and thus produces multiple trace ID trees, the algorithm computes the metrics for each tree separately and outputs the mean, minimum, maximum, and standard deviation for each metric.

All metrics depend on the size of leaves in the trace ID tree. For $n$ traces, let $T=\{0,1,\ldots,n-1\}$ be the set of trace IDs. The set of leaves $L$ for a given trace ID tree is then defined as $L=\{L_i\,|\,L_i\subseteq T \land L_i\neq \emptyset\}$ with $L_1\cup L_2\cup\ldots\cup L_\ell=T$ and $L_i\cap L_j=\emptyset$ for $L_i,L_j\in T$ and $i\neq j$. This can be read as the tree having $\ell$ leaves $L_i$ ($i=1,\ldots,\ell$), where each $L_i$ holds the trace IDs ending up in this particular leaf. Those traces are considered identical.

Let $X\colon T\rightarrow\mathbb{N}$ be a random variable for picking a trace ID. The trace IDs are uniformly distributed, hence $\Pr[X=k]=\frac{1}{n}$ for each $k=1,\ldots,n$. Let $Y\colon L\rightarrow \mathbb{N}$ be a random variable for observing a particular trace, with $\Pr[Y=i]=\frac{|L_i|}{|T|}=\frac{|L_i|}{n}$ for $i=1,\ldots,\ell$.

\bheading{Mutual information} measures the \emph{average amount of information} an attacker learns when observing a trace.

The MI of the trace ID $X$ and the observed trace $Y$ is defined as
\begin{align*}
    I(X,Y)&=\sum_{k=1}^{|T|}\sum_{i=1}^{|L|}\Pr[X=k,Y=i]\cdot\log_2\left(\frac{\Pr[X=k,Y=i]}{\Pr[X=k]\cdot\Pr[Y=i]}\right).
\end{align*}
With
\begin{equation*}
    \Pr[X=k,Y=i]=\begin{cases}
        \hfil 0 & \text{if}\ k\notin L_i\\
        \frac{1}{n} & \text{if}\ k\in L_i
    \end{cases}
\end{equation*}
we get
\begin{equation*}
    I(X,Y)=\sum_{i=1}^{|L|}\frac{|L_i|}{n}\cdot\log_2\left(\frac{\frac{1}{n}}{\frac{1}{n}\cdot\frac{|L_i|}{n}}\right)=\frac{1}{n}\sum_{i=1}^{\ell}|L_i|\cdot\log_2\left(\frac{n}{|L_i|}\right).
\end{equation*}

The value of $I(X,Y)$ can be interpreted as bits: In the best case, there is only one leaf containing all trace IDs, such that the attacker learns nothing (0 bits). In the worst case, with one leaf for each trace ID, the attacker learns $\log_2(n)$ bits. The MI of the example in Figure~\ref{fig:trace-id-tree} is $\frac{1}{6}\left(2\cdot 2\cdot\log_2(3)+2\cdot 1\cdot\log_2(6)\right)\approx 1.33$ bits.

This metric has a few drawbacks: Due to its logarithmic nature, with an increasing number of traces it only grows slowly. Another shortcoming is the averaging, i.e., a high leakage in a few cases may get suppressed by the smaller leakage of all other cases. Finally, it may be mistakenly interpreted as additive due to its ``bits'' unit (i.e., 10 instructions leaking 3 bits each does not mean that there is a leakage of 30 bits). However, it does perform well for small and balanced leakages, e.g., when an instruction constantly divides the traces into two groups of similar size.

\bheading{Conditional guessing entropy} measures the \emph{expected number of guesses} an attacker needs for associating a given trace with a secret input.
The conditional GE $G(X\,|\,Y)$ for determining a trace ID, modeled as random variable $X$, for a known value of an observed trace (random variable $Y$) is calculated as 
\begin{align}
    G(X\,|\,Y)&=\sum_{i=1}^{|L|}\Pr[Y=i]\cdot G(X\,|\,Y=i) \nonumber\\
    &=\sum_{i=1}^{|L|}\Pr[Y=i]\cdot\sum_{k=1}^{|T|}k\cdot \Pr[X=k\,|\,Y=i]. \label{eq:conditional-guessing-entropy}
\end{align}
Since
\begin{equation*}
    \Pr[X=k\,|\,Y=i]=\begin{cases}
        \hfil 0 & \text{if}\ k\notin L_i\\
        \frac{1}{|L_i|} & \text{if}\ k\in L_i,
    \end{cases}
\end{equation*}
we can simplify \eqref{eq:conditional-guessing-entropy} to
\begin{align*}
    G(X\,|\,Y)&=\sum_{i=1}^{\ell}\Pr[Y=i]\cdot\frac{1}{|L_i|}\sum_{k=1}^{|L_i|}k %
    =\frac{1}{2n}\sum_{i=1}^{\ell}|L_i|\cdot(|L_i|+1).
\end{align*}
Note that the value of $G(X\,|\,Y)$ is upper-bounded by $\frac{n+1}{2}$, which is the best case where there is only one leaf which contains all trace IDs, i.e., all traces are identical. For the example in Figure~\ref{fig:trace-id-tree}, we get $G(X\,|\,Y)=\frac{1}{2\cdot 6}(6+2+6+2)\approx 1.33$ guesses.

Small values for the conditional GE convey that an instruction sequence leads to almost unique traces, implying that there is widespread leakage affecting most to all traces. On the other side, a high value means that most traces are similar and do not leak much information. However, this being an average measure just like MI, there may well be special cases where there is a very high leakage. Those risk being obscured by this metric, thus we add an additional worst-case metric designed for catching these cases.

\bheading{Minimal conditional guessing entropy} measures the \emph{minimal number of guesses} an attacker needs for associating a given trace with a secret input.
It is calculated similarly to the conditional GE, but takes the minimum of all individual outcomes instead of weighting them:
\begin{equation*}
    \hat{G}(X\,|\,Y)=\min_{i=1,\ldots,|L|} G(X\,|\,Y=i)=\min_{i=1,\ldots,\ell} \frac{|L_i|+1}{2}.
\end{equation*}
For the example in Figure~\ref{fig:trace-id-tree}, we get $\hat{G}(X\,|\,Y)=\min\{1.5,1,1.5,1\}=1$ guesses, i.e., there is at least one trace that is unique.

Minimal GE is the most definite leakage measure; it gives the number of guesses needed for the trace which leaks most. A high value for the minimal GE affirms that there is no outlier with high leakage. We thus recommend using this metric when evaluating the severity of a detected leakage.

\subsubsection{Leakage severity and score}
\label{sec:leakage_score}
While the full analysis report provides detailed information about each leakage, we also seek to condense this information into a single, uniform score, such that the developer can quickly prioritize.
That score should require little context: The developer should not need to be familiar with entropy, nor know analysis details like the particular number of test cases, which determines the upper bounds for the various metrics. 
Additionally, providing a single score allows easy integration of the leakage report into the user interface of modern development platforms like GitLab. The platform can then use that score for sorting and assigning a severity to the leakage.

We chose minimal GE for computing the leakage score, as it represents the worst-case leakage. 
Instead of reporting the minimal GE value directly, we map it onto a linear scale of 0 to 100, where 0 corresponds to a minimal GE of $\frac{n+1}{2}$ (i.e., no leakage), and 100 corresponds to a minimal GE of $1$ (i.e., maximum leakage). 
If there are multiple trace ID trees for a given instruction (see Section~\ref{sec:leakage-analysis:leakage-analysis:building-trace-id-trees}), we show the mean and the standard deviation over the individual minimal GE values. 

\subsection{Implementation}
We implemented the described algorithm as a new analysis module in \mwci/'s source tree. It integrates directly into the leakage analysis pipeline, i.e., it receives and handles preprocessed traces from the previous pipeline stage, the trace preprocessor.
The tree is implemented as a recursive data structure, where each node holds a list of successor and split nodes. We do not store the consecutive non-diverging trace entries as a plain \texttt{ITraceEntry} list (as is suggested in the algorithm description), but as full-featured tree nodes as well. Apart from making the code more readable, this simplifies adding new divergences and storing temporary data for the final leakage analysis step, at the cost of additional memory overhead (we discuss this trade-off in Section~\ref{sec:discussion:performance}).

Our implementation offers functionality for generating leakage reports and other detailed analysis result files optimized for readability, including an optional full call tree dump for debugging purposes. All features can be controlled via the \mwci/ configuration file infrastructure, allowing easy adoption of the new analysis module. In total, the module has 1,363 lines of C\# code.

\section{\javascript/ Leakage Analysis}
\label{sec:javascript-leakage-analysis}

We now show how we can apply \mwci/'s generic analysis methods to \javascript/ libraries, despite them being originally designed for binary analysis. First, we present a simple trace generator relying on the \jalangitwo/ instrumentation framework. Then, we show how these traces can be preprocessed such that they use the generic trace format from \mwci/.

\subsection{Instrumenting \javascript/ code}
\label{sec:javascript-leakage-analysis:instrumenting}

\mwci/ expects multiple execution traces with varying secret input for the analyzed target function. These execution traces are then fed into various analysis modules for finding non-constant-time behavior, i.e. \controlflow/ or \dataflow/-dependencies from secret input.
A trace needs to contain the following information:
\begin{itemize}
    \item Address and size of all loaded program modules (e.g., binaries or source files, called ``images'' internally);
    \item the control-flow of the analyzed program, encoded as a sequence of branch source and target addresses;
    \item address and size of all heap memory objects; and
    \item the instructions and target addresses of all memory accesses.
\end{itemize}

We translate this to \javascript/ by collecting a trace of all executed code lines, and recording access offsets to any object or array. For instrumentation, we use \jalangitwo/~\cite{jalangi2github}. \jalangitwo/ instruments the code at load time by inserting callbacks before and after certain source tokens, e.g., conditionals, expressions or return statements.

First, we register the provided \emph{SMemory} analysis module, which assigns a shadow object to each object, that contains a unique ID and the object value, allowing us to map accesses to known objects.
We then create an own analysis front-end, called \emph{tracer}, which registers some callbacks to record the necessary information and write it to a file for further processing. The tracer has 252 lines of code, and is chained after the \emph{SMemory} analysis, which supplies the means for memory access tracking.

\subsection{Trace File Structure}
\label{sec:javascript-leakage-analysis:trace-file-structure}

\begin{figure}[t]
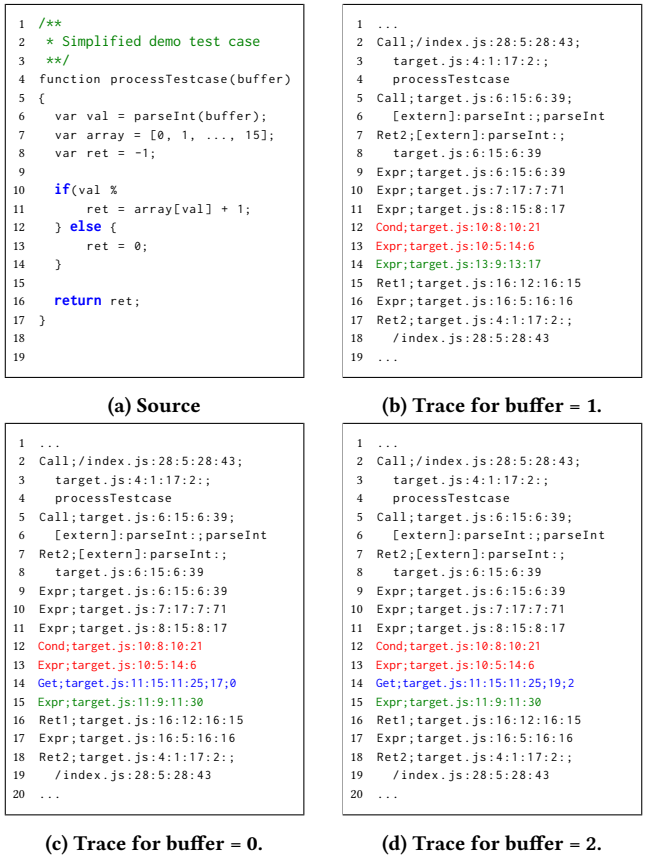

    \centering
    \begin{subfigure}[b]{0.47\columnwidth}
        \centering
        \begin{mdframed}[style=listingframe]
        \begin{lstlisting}[language=C,basicstyle=\tiny\ttfamily,numbers=left,xleftmargin=1ex,numbersep=5pt,showlines=true]
/** 
 * Simplified demo test case 
 **/
function processTestcase(buffer)
{
  var val = parseInt(buffer);
  var array = [0, 1, ..., 15];
  var ret = -1;

  if(val %
      ret = array[val] + 1;
  } else {
      ret = 0;
  }

  return ret;
}
 
 
        \end{lstlisting}
        \end{mdframed}
        \caption{Source}
        \label{fig:jstrace:source}
    \end{subfigure}\hfill%
    \begin{subfigure}[b]{0.47\columnwidth}
        \centering
        \begin{mdframed}[style=listingframe]
        \begin{lstlisting}[basicstyle=\tiny\ttfamily,numbers=left,xleftmargin=1ex,numbersep=5pt,escapeinside={<@}{@>}]
...
Call;/index.js:28:5:28:43;
  target.js:4:1:17:2:;
  processTestcase
Call;target.js:6:15:6:39;
  [extern]:parseInt:;parseInt
Ret2;[extern]:parseInt:;
  target.js:6:15:6:39
Expr;target.js:6:15:6:39
Expr;target.js:7:17:7:71
Expr;target.js:8:15:8:17
<@\textcolor{red}{Cond;target.js:10:8:10:21}@>
<@\textcolor{red}{Expr;target.js:10:5:14:6}@>
<@\textcolor{halfgreen}{Expr;target.js:13:9:13:17}@>
Ret1;target.js:16:12:16:15
Expr;target.js:16:5:16:16
Ret2;target.js:4:1:17:2:;
  /index.js:28:5:28:43
...
        \end{lstlisting}
        \end{mdframed}
        \caption{Trace for buffer = 1.}
        \label{fig:jstrace:trace1}
    \end{subfigure}
    \begin{subfigure}[b]{0.47\columnwidth}
        \centering
        \begin{mdframed}[style=listingframe]
        \begin{lstlisting}[basicstyle=\tiny\ttfamily,numbers=left,xleftmargin=1ex,numbersep=5pt,escapeinside={<@}{@>}]
...
Call;/index.js:28:5:28:43;
  target.js:4:1:17:2:;
  processTestcase
Call;target.js:6:15:6:39;
  [extern]:parseInt:;parseInt
Ret2;[extern]:parseInt:;
  target.js:6:15:6:39
Expr;target.js:6:15:6:39
Expr;target.js:7:17:7:71
Expr;target.js:8:15:8:17
<@\textcolor{red}{Cond;target.js:10:8:10:21}@>
<@\textcolor{red}{Expr;target.js:10:5:14:6}@>
<@\textcolor{blue}{Get;target.js:11:15:11:25;17;0}@>
<@\textcolor{halfgreen}{Expr;target.js:11:9:11:30}@>
Ret1;target.js:16:12:16:15
Expr;target.js:16:5:16:16
Ret2;target.js:4:1:17:2:;
  /index.js:28:5:28:43
...
        \end{lstlisting}
        \end{mdframed}
        \caption{Trace for buffer = 0.}
        \label{fig:jstrace:trace0}
    \end{subfigure}\hfill%
    \begin{subfigure}[b]{0.47\columnwidth}
        \centering
        \begin{mdframed}[style=listingframe]
        \begin{lstlisting}[basicstyle=\tiny\ttfamily,numbers=left,xleftmargin=1ex,numbersep=5pt,escapeinside={<@}{@>}]
...
Call;/index.js:28:5:28:43;
  target.js:4:1:17:2:;
  processTestcase
Call;target.js:6:15:6:39;
  [extern]:parseInt:;parseInt
Ret2;[extern]:parseInt:;
  target.js:6:15:6:39
Expr;target.js:6:15:6:39
Expr;target.js:7:17:7:71
Expr;target.js:8:15:8:17
<@\textcolor{red}{Cond;target.js:10:8:10:21}@>
<@\textcolor{red}{Expr;target.js:10:5:14:6}@>
<@\textcolor{blue}{Get;target.js:11:15:11:25;19;2}@>
<@\textcolor{halfgreen}{Expr;target.js:11:9:11:30}@>
Ret1;target.js:16:12:16:15
Expr;target.js:16:5:16:16
Ret2;target.js:4:1:17:2:;
  /index.js:28:5:28:43
...
        \end{lstlisting}
        \end{mdframed}
        \caption{Trace for buffer = 2.}
        \label{fig:jstrace:trace2}
    \end{subfigure}
    \caption{Traces created by \mwci/ for a \javascript/ toy example. Indented lines are wrapped for readability and are formatted in a single line in the original trace file.}
    \label{fig:jstrace}
\end{figure}

Figure \ref{fig:jstrace} illustrates the structure of the trace files for a simple toy example. The example has an input-dependent branch in line 10 and a secret-dependent memory access in line 11, which should be detected by our analysis toolchain.

Each trace is structured as follows: The first element defines the type of the trace event, e.g. \textit{Call} or \textit{Expr} (for Expression). This is followed by the exact source location of the event, meaning the script file name with start/end line and column number.
For a \textit{Call}, the first location describing the source of the call is followed by a second location describing the target, which in turn is followed by the name of the called function. \textit{Expr} entries log the locations of all executed expressions; this information is only needed for reconstructing control flow edges. Similarly, \textit{Ret1} records the occurrence of a \texttt{return}-statement, which must be tracked due to not being covered by an expression. \textit{Ret2} is generated \emph{after} a function has returned, and records the entire ranges of the function call and the executed function; however, the associated callback does not know where the control flow originated from, thus the necessity of tracking expressions and \texttt{return} statements. The same is true for \textit{Cond} entries, which mark the execution of a conditional and thus the begin of a control-flow edge.

To illustrate this, Figure \ref{fig:jstrace:trace1} shows the case of a taken \texttt{else}-branch with the assignment \texttt{ret = 0} in line 14 of the trace. If we compare this trace to the Figures \ref{fig:jstrace:trace0} and \ref{fig:jstrace:trace2}, which both show a taken \texttt{if}-branch, it becomes apparent that the control-flow deviation only shows up due to the differences in lines 14 and 15; everything else is identical. Thus, only tracking all expressions and read/write operations allows us to reconstruct the entire control flow.

Comparing line 14 of Figures~\ref{fig:jstrace:trace0} and~\ref{fig:jstrace:trace2} demonstrates how the traces enable us to discover secret-dependent memory accesses. The last two elements of the \textit{Get} entry represent the ID of the shadow object, and the accessed property or offset. Both elements differ between the traces: The object IDs are assigned by \jalangitwo/ and thus vary for subsequent invocations of \texttt{processTestcase}, and the accessed offset depends on the input. The analysis conducted by \mwci/ will later match the object IDs belonging to identical objects between traces, such that it can compare the offsets.

This example shows a very short excerpt of a trace for a toy program. Analyzing real world code may result in traces with millions of events, resulting in huge files. To reduce the storage overhead, we compress the trace by shortening strings and encoding repeating lines. For most targets, these measures are sufficient to keep the trace files within a few ten megabytes. Additional compression could be achieved e.g. by using LZMA, which due to the high rate of repetitions and hence low entropy usually manages to bring down the trace file size to a few hundred kilobytes.

\subsection{Trace Preprocessing}
These raw traces are not yet suitable for use by \mwci/; we need to translate the sequence of executed lines to branch entries, generate allocation information for the objects showing up in the traces, and finally produce compatible binary traces, which can be fed into analysis modules like the one described in Section~\ref{sec:leakage-analysis}.
For this, we implemented a new preprocessor module, which has 702 lines of code and resides in a plugin. The module iterates through each entry of the raw trace, generating a preprocessed trace on-the-fly. It recognizes branches by waiting for the next code location that is outside the corresponding conditional; if an access to a previously unknown object is detected, an allocation is created.

Note that our analysis module is designed for binary analysis, i.e., it works with actual memory addresses and offsets. In fact, this proves valuable for later analysis, as this simplifies encoding trace entries and gives clear identifiers for referring to certain instructions. Thus, we chose to generate a mapping of observed source locations to dummy addresses, by encoding the line and column numbers onto a base address belonging to the respective source file. This mapping is stored in a special map file, such that it can be mapped back to a human-readable source line after analysis.

In summary, we now have a tool chain that instruments \javascript/ programs, generates raw execution traces and converts them into the \mwci/ binary trace format, allowing us to analyze arbitrary \javascript/ software with the existing and new generic analysis algorithms, without having to create a dedicated analysis tool.

\section{Integration into Development Workflow}
\label{sec:ci}

In this section, we show how one can simplify usage of \mwci/ to a degree that it only needs a one-time effort by the developer to set it up and register the functions that need to be analyzed. From that point, the tool is part of the CI pipeline of the respective library, and runs each time a new commit is submitted. The developer is then able to easily verify whether a code change introduces new leakages, without requiring any manual intervention.

\subsection{Dockerizing the Analysis Framework}
\label{sec:ci:dockerization}
In order to use the analysis framework in an automated environment, we must ensure that all its dependencies are present and the environment is configured correctly. 
For this task, common CI systems allow the use of Docker containers. When a job starts, a new container is started from a predefined Docker image.
The CI system checks out the current source code and then executes a user-defined script within the container. This has the advantage of being independent of the host system:  %
The analysis job may run on the developer's private server, but also on cloud infrastructure administrated by external providers.
We thus create a pre-configured Docker image containing the components needed for our \javascript/ analysis: The \jalangitwo/ runtime, the analysis script and the \mwci/ binaries.
The image is uploaded to a Docker registry. %

\subsection{Analysis Template}
\label{sec:ci:template}
Having solved the installation and configuration problem, we now need to setup the necessary infrastructure to actually run the analysis for the specific library. Instead of requiring the developer to dive into the proper usage of the analysis toolchain, we designed a template that is simple and generic enough to work with most libraries, and which only needs minimal understanding and adjustment. The resulting file structure is depicted in Figure~\ref{fig:ci-file-structure}.

\begin{figure}[t]
\begin{mdframed}[style=listingframe]
\begin{lstlisting}[basicstyle=\linespread{1.1}\scriptsize\ttfamily,language=]
/                           (*@\color{halfgreen}{\# Project source tree}@*)
 (*@\noindentrule@*) index.js                  (*@\color{halfgreen}{\# Analysis entrypoint}@*)
 (*@\noindentrule@*) package.json
 (*@\noindentrule@*) ...
 (*@\noindentrule@*) microwalk/                (*@\color{halfgreen}{\# Analysis-specific files}@*)
 (*@\noindentrule@*)  (*@\noindentrule@*) analyze.sh              (*@\color{halfgreen}{\# Script executed by CI}@*)
 (*@\noindentrule@*)  (*@\noindentrule@*) config-preprocess.yml   (*@\color{halfgreen}{\# Microwalk config. for preprocessing}@*)
 (*@\noindentrule@*)  (*@\noindentrule@*) config-analyze.yml      (*@\color{halfgreen}{\# Microwalk config. for analysis}@*)
 (*@\noindentrule@*)  (*@\noindentrule@*) target-aes.js           (*@\color{halfgreen}{\# AES target}@*)
 (*@\noindentrule@*)  (*@\noindentrule@*) target-rsa.js           (*@\color{halfgreen}{\# RSA target}@*)
 (*@\noindentrule@*)  (*@\noindentrule@*) ...
 (*@\noindentrule@*)  (*@\noindentrule@*) testcases/              (*@\color{halfgreen}{\# Input files for trace generation}@*)
 (*@\noindentrule@*)  (*@\noindentrule@*)  (*@\noindentrule@*) target-aes/
 (*@\noindentrule@*)  (*@\noindentrule@*)  (*@\noindentrule@*)  (*@\noindentrule@*) 0.testcase
 (*@\noindentrule@*)  (*@\noindentrule@*)  (*@\noindentrule@*)  (*@\noindentrule@*) 1.testcase
 (*@\noindentrule@*)  (*@\noindentrule@*)  (*@\noindentrule@*)  (*@\noindentrule@*) ...
 (*@\noindentrule@*)  (*@\noindentrule@*)  (*@\noindentrule@*) target-rsa/
 (*@\noindentrule@*)  (*@\noindentrule@*)  (*@\noindentrule@*)  (*@\noindentrule@*) 0.testcase
 (*@\noindentrule@*)  (*@\noindentrule@*)  (*@\noindentrule@*)  (*@\noindentrule@*) 1.testcase
 (*@\noindentrule@*)  (*@\noindentrule@*)  (*@\noindentrule@*)  (*@\noindentrule@*) ...
 (*@\noindentrule@*)  (*@\noindentrule@*)  (*@\noindentrule@*) ...
\end{lstlisting}
\end{mdframed}
\caption{Generic source tree of a \javascript/ project containing our analysis template.}
\label{fig:ci-file-structure}
\end{figure}

The template features a script file \texttt{index.js}, which serves as analysis entry point and is responsible for loading test cases and executing the target implementations. A \emph{target} is any independently testable code unit, e.g., a single primitive in a cryptographic library. The individual \texttt{microwalk/target-*.js} script files consist of a single function, that receives the current test case data buffer and is expected to call the associated library code. Each target also needs a number of test cases, which may have a custom format and thus need to be generated once by the developer. The test cases are stored in the \texttt{microwalk/testcases/} subdirectory.
Finally, the \texttt{microwalk} folder has a bash script \texttt{analyze.sh}, that is called by the CI. The analysis script iterates through the target files, and runs the \mwci/ pipeline. The \mwci/ configuration is located in two generic YAML files, which can be adjusted by the developer if they wish to use other analysis modules or options than the preconfigured ones.

The abstractions offered by our template allows the developer to focus on supplying simple wrappers for their library interface and generating a number of random test cases; everything else is taken care of by the existing scripts. We implemented a similar template for compiled software, so the approach is the same for C libraries.

\subsection{Reports}
\label{sec:ci:reports}
After the CI job has completed, it yields a couple of analysis result files. As of our analysis objectives in Section~\ref{sec:leakage-analysis}, these files are designed to be human-readable and offer as much insight into a leakage as possible. However, if there are a lot of leakage candidates, going through this list may be tedious, especially if the result files are stored separately and need to be inspected manually for each commit. We thus looked into ways for integrating these results into the usual development workflow.

\begin{figure}
    \begin{tikzpicture}
        \node[draw,align=left,text width=23em,minimum height=3.3em] (box1) at (1,0) {
            \scriptsize
            \textbf{Critical} - (target-toy-example) Found vulnerable
            memory access instruction, leakage score 100.00\% +/- 0\%. 
            Check analysis result in artifacts for details.\\
            in \textcolor{halfblue}{target.js:11}
        };
        \node[draw,align=left,text width=0.8em,left=0em of box1,minimum height=3.3em, text depth=1.8em] (icon1) {
            \includegraphics[width=0.8em]{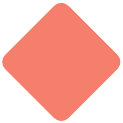}
        };
        \node[draw,align=left,below=0em of box1,text width=23em,minimum height=3.3em] (box2) {
            \scriptsize
            \textbf{Major} - (target-toy-example) Found vulnerable
            jump instruction, leakage score 53.33\% +/- 0\%.
            Check analysis result in artifacts for details.\\
            in \textcolor{halfblue}{target.js:10}
        };
        \node[draw,text width=0.8em,left=0em of box2,minimum height=3.3em,text depth=1.8em] (icon2) {
            \includegraphics[width=0.8em]{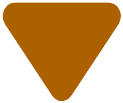}
        };
    \end{tikzpicture}
    \caption{GitLab report for the toy example from Figure~\ref{fig:jstrace:source}. The leakage score is a relative representation of the minimal GE as explained in Section \ref{sec:leakage_score}.}
    \label{fig:gitlab-example-report}
\end{figure}

For GitLab, there is a \emph{Code Quality Reports}~\cite{gitlab-code-quality} feature, which shows up in the merge request UI.
It allows to assign a severity, a description and a source code file and line to each entry, which makes it suitable to display the results from our leakage analysis.
\mwci/ consolidates the analysis result into a report that can be parsed by GitLab. For this, the leakages must be mapped to their originating locations in the source code. This is straightforward for \javascript/, as this information already shows up in the analysis result file; for binary programs, we resort to parsing the DWARF debug information in order to map offsets to file names and lines. The code quality report also shows a severity of a given problem, which can be one of \emph{info}, \emph{minor}, \emph{major}, \emph{critical} and \emph{blocker} (a continuous scale is not supported). Assigning these levels to specific leakages is somewhat arbitrary and depends on the preferences of the individual developer; we settled for \emph{minor} if the minimal GE is higher than 80\% of its upper bound, \emph{critical} if the minimal GE is lower than 20\% of its upper bound, and \emph{major} for everything in between. This ensures that instances with high leakage are displayed prominently.
Figure~\ref{fig:gitlab-example-report} shows an example report. %

\section{Evaluation and Discussion}
\label{sec:discussion}
To evaluate \mwci/, we applied it to several popular \javascript/ crypto libraries. In the following, we describe our experimental setup and discuss the performance and discovered vulnerabilities.

\subsection{Experimental Setup}
As targets we pulled eight popular
\javascript/ libraries for cryptography and utility functions from \npm/, and set up a local GitLab repository for each. Using the version from \npm/ instead of the version from GitHub allows us to analyze the code deployed to millions of users. 
We then applied our template and created \texttt{target-*.js} files for selected cryptographic primitives and utility functions that deal with secret data. For each target we generated 16 random test cases, which were subsequently checked in into the source tree. 

The GitLab instance takes care of managing the CI jobs and visualizing the resulting code quality reports. The analysis jobs themselves are executed through a Docker-based GitLab Runner on a separate machine (build server), which has an AMD EPYC 7763 processor with 128 GB DDR4 RAM. We configured the \mwci/ trace preprocessor step to use up to 4 CPU cores.
After all CI jobs had completed, we collected the performance statistics generated by GitLab and the CI jobs, and went through the leakage reports. The results are visualized in Table~\ref{tab:js-mw-libs}.

\begin{table*}[t]
    \caption{Targets analyzed with \javascript/ \mwci/, performance metrics and the number of detected leakages (total and unique code lines). ``Tr. CPU'' shows the CPU time for generating the raw traces, ``Prep. CPU'' for trace preprocessing, and ``An. CPU'' for the analysis step. ``Duration'' denotes the wall clock time spent for the entire CI job (including setup and cleanup). Finally, ``Prep. RAM'' and ``An. RAM`` show the peak memory usage for the preprocessing and analysis steps, respectively.}
    \label{tab:js-mw-libs}
    \centering
    \begin{tabular}{ l l r r r r r r r r }
        Target & Type & Tr. CPU & Prep. CPU & An. CPU & Duration & Prep. RAM & An. RAM & \# Leakages & \# Unique \\ \hline \hline
        
        \multicolumn{4}{l}{\textbf{aes-js}~\cite{target-aes-js} 3.1.2, $\approx$ 800k weekly downloads} & & & & \\
            \quad AES-ECB & cipher & 1 sec & $<$ 1 sec & $<$ 1 sec & 7 sec & 294 MB & 180 MB & 16 & 16 \\
        \hline
        
        \multicolumn{4}{l}{\textbf{base64-js}~\cite{target-base64-js} 1.5.1, $\approx$ 28M weekly downloads} & & & & \\
            \quad base64-encode & utility & $<$ 1 sec & $<$ 1 sec & $<$ 1 sec & 6 sec & 283 MB & 173 MB & 7 & 7 \\
            \quad base64-decode & utility & $<$ 1 sec & $<$ 1 sec & $<$ 1 sec & 6 sec & 291 MB & 189 MB & 7 & 7 \\
        \hline
        
        \multicolumn{4}{l}{\textbf{crypto-js}~\cite{target-crypto-js} 4.1.1, $\approx$ 4M weekly downloads} & & & & \\
            \quad AES-ECB & cipher & 2 sec & $<$ 1 sec & $<$ 1 sec & 8 sec & 289 MB & 191 MB & 44 & 44 \\
            \quad Rabbit & cipher & 2 sec & $<$ 1 sec & $<$ 1 sec & 8 sec & 304 MB & 182 MB & 0 & 0 \\
            \quad base64-encode & utility & 4 sec & 1 sec & 1 sec & 10 sec & 349 MB & 250 MB & 0 & 0 \\
            \quad base64-decode & utility & 3 sec & $<$ 1 sec & $<$ 1 sec & 9 sec & 339 MB & 225 MB & 2 & 2 \\
            \quad pbkdf2 & utility & 5 sec & 1 sec & 1 sec & 11 sec & 384 MB & 221 MB & 0 & 0 \\
        \hline
        
        \multicolumn{4}{l}{\textbf{elliptic}~\cite{target-elliptic} 6.5.4, $\approx$ 13M weekly downloads} & & & & \\
            \quad secp256k1 & signature & 110 sec & 26 sec & 21 sec & 139 sec & 853 MB & 3,123 MB & 58 & 45 \\
            \quad p192 & signature & 237 sec & 35 sec & 12 sec & 261 sec & 2,112 MB & 1,835 MB & 98 & 57 \\
            \quad p224 & signature & 303 sec & 45 sec & 14 sec & 334 sec & 1,700 MB & 2,357 MB & 76 & 58 \\
            \quad p256 & signature & 469 sec & 84 sec & 45 sec & 545 sec & 3,347 MB & 6,674 MB & 78 & 50 \\
            \quad p384 & signature & 977 sec & 145 sec & 45 sec & 1,063 sec & 3,383 MB & 7,522 MB & 391 & 53 \\
            \quad ed25519 & signature & 175 sec & 46 sec & 32 sec & 222 sec & 2,884 MB & 4,607 MB & 111 & 40 \\
        \hline
        
        \multicolumn{4}{l}{\textbf{js-base64}~\cite{target-js-base64} 3.7.2, $\approx$ 6M weekly downloads} & & & & \\
            \quad base64-encode & utility & $<$ 1 sec & $<$ 1 sec & $<$ 1 sec & 6 sec & 290 MB & 187 MB & 0 & 0 \\
            \quad base64-decode & utility & $<$ 1 sec & $<$ 1 sec & $<$ 1 sec & 6 sec & 290 MB & 155 MB & 0 & 0 \\
        \hline
        
        \multicolumn{4}{l}{\textbf{node-forge}~\cite{target-node-forge} 1.2.1, $\approx$ 17M weekly downloads} & & & & \\
            \quad AES-ECB & cipher & 5 sec & $<$ 1 sec & $<$ 1 sec & 11 sec & 298 MB & 193 MB & 36 & 36 \\
            \quad AES-GCM & cipher & 9 sec & 2 sec & 2 sec & 16 sec & 387 MB & 349 MB & 126 & 52 \\
            \quad base64-encode & utility & 5 sec & $<$ 1 sec & $<$ 1 sec & 11 sec & 287 MB & 192 MB & 0 & 0 \\
            \quad base64-decode & utility & 5 sec & $<$ 1 sec & $<$ 1 sec & 11 sec & 296 MB & 198 MB & 4 & 4 \\
            \quad rsa & signature & 62 sec & 18 sec & 13 sec & 82 sec & 364 MB & 1,926 MB & 223 & 111 \\
            \quad ed25519 & signature & 124 sec & 33 sec & 9 sec & 144 sec & 1,145 MB & 509 MB & 0 & 0 \\
        \hline
        
        \multicolumn{4}{l}{\textbf{pbkdf2}~\cite{target-pbkdf2} 3.1.2, $\approx$ 13M weekly downloads} & & & & \\
            \quad pbkdf2 & utility & $<$ 1 sec & $<$ 1 sec & $<$ 1 sec & 6 sec & 298 MB & 179 MB & 0 & 0 \\
        \hline
        
        \multicolumn{4}{l}{\textbf{tweetnacl}~\cite{target-tweetnacl} 1.0.3, $\approx$ 21M weekly downloads} & & & & \\
            \quad secretbox & cipher & 2 sec & $<$ 1 sec & $<$ 1 sec & 8 sec & 288 MB & 189 MB & 0 & 0 \\
            \quad box & asymmetric & 75 sec & 20 sec & 7 sec & 91 sec & 335 MB & 487 MB & 0 & 0 \\
            \quad ed25519 & signature & 117 sec & 33 sec & 9 sec & 138 sec & 1,137 MB & 509 MB & 0 & 0 \\
        \hline

    \end{tabular}
\end{table*}

\subsection{Performance}
\label{sec:discussion:performance}

\subsubsection{Computation time}
\label{sec:discussion:performance:time}
The CPU time spent for trace generation, preprocessing and analysis mostly depends on two factors. First, it correlates with the complexity of the analyzed targets: For the investigated libraries, symmetric algorithms and utility functions performed very well, while asymmetric primitives took significantly longer, which is expected. Second, the CPU time scales linearly with the number of test cases. We discuss the corresponding trade-off between accuracy and performance in Section~\ref{sec:discussion:testcases}.

The computational cost for the \textbf{trace generation step} mainly stems from the instrumentation itself, as our tracer script is already quite minimal. Significant optimizations would thus need to target the \jalangitwo/ implementation.
The CPU time spent for the \textbf{preprocessing step} correlates with the size of the raw traces. The implementation is parallelized, so each trace can be processed independently.
Profiling shows a slight bottleneck in the string parsing code, so switching to a binary trace format may further improve preprocessing performance, at the cost of higher code complexity in the trace generation.
The \textbf{analysis step} took less than one CPU minute for every investigated target; this underlines the efficiency of the presented analysis algorithm, and that it is fast enough to be used in a productive setting. The time spent for the analysis mostly depends on the trace size, as when building the call tree, each trace entry is converted into a tree node or embedded into an existing one. Another factor is the number of leakages, as is apparent when comparing the analysis times of the various \texttt{ed25519} implementations. 
The measured \textbf{overall duration} heavily depends on where most CPU time is spent: While the trace generation and the analysis are mostly sequential, the trace preprocessing is heavily parallelized. Thus, a high CPU time for preprocessing does contribute less to the overall duration.
Apart from one outlier, \texttt{elliptic}'s \texttt{p384}, the measured times stayed well within a few minutes, which can be considered acceptable for productive use in a CI pipeline.

\subsubsection{Memory usage}
\label{sec:discussion:performance:memory}
The inherently different pipeline steps also reflect in different memory requirements.
The \textbf{trace generation step} has a negligible memory footprint, which mostly depends on the size of the array that is used for buffering trace entries before writing them to the output file.

The memory consumption of the \textbf{preprocessing step} is mainly caused by loading chunks of the trace file into memory and decompressing them. Parallelization of the preprocessing step means that several trace files are being held in memory simultaneously.
The memory usage of the preprocessing can be reduced by decreasing the number of parallel threads (4 in our experiment).

In the call tree \textbf{analysis step}, the memory demand is driven by the size of the preprocessed traces and, most notably, their level of divergence. If the target is constant-time and thus all traces are identical, the tree does not have any split nodes, so all traces end up in the same nodes. Adding a trace ID to an existing node does not involve any significant memory cost, as the trace IDs assigned to a call tree node are stored as a bitfield. 

However, if the traces heavily diverge, the analysis produces many split nodes with partially redundant subtrees. This distinction becomes apparent by the implementations of \texttt{ed25519} in \texttt{elliptic} and in \texttt{tweetnacl}: While using comparable tracing and preprocessing time, the constant-time implementation in \texttt{tweetnacl} requires much less memory than the implementation in \texttt{elliptic}, which relies on the leaking \texttt{bn.js} and \texttt{hash.js} libraries. 
Through continuously applying \mwci/ and mitigating non-constant-time behavior such that only small leakages pop up during analysis, the peak memory usage of the analysis step can be kept within the bounds of a typical CI environment.
Overall, the \textbf{peak memory usage} of \mwci/ is on an acceptable level.
The highest memory consumption was observed when analyzing \texttt{elliptic}'s \texttt{p384}. This is certainly a worst case example, as large parts of its code are non-constant time, while \mwci/ is optimized for finding mid-level leakages in an otherwise fairly constant-time software.
However, most of \texttt{p384}'s code is shared with the other curve implementations, which contain the same leakages, but can be analyzed more efficiently. Also, a significant part of the identified leakages reside in the the SHA-512 implementation of \texttt{hash.js}, which should be analyzed separately.

As expected, more complex algorithms like asymmetric cryptography require more memory in the analysis. But, even those only require an amount of memory which, today, is commonly available.

\subsection{Vulnerabilities}
\label{sec:discussion:vulnerabilites}
Our leakage analysis identified many leakages in the given libraries. We evaluated whether those are in fact actual vulnerabilities, and discuss a few examples in the following. In general, the leakages were correctly assigned to the respective leaking code lines, and we did not encounter any false positives (i.e., code lines that don't leak by themselves).
In addition to the report shown in the user interface (Figure~\ref{fig:gitlab-example-report}), a detailed leakage report is generated, which provides the full calling context for each leakage and shows how the different test cases contributed to tree divergences. %

\subsubsection{Leakages in AES}
All investigated implementations of AES use table lookups into S-boxes or precomputed T-tables, making those highly susceptible to timing attacks. 
The exploitability of such lookups was previously shown in other work~\cite{bernstein2005cache}. All leakages found in \texttt{aes-js} by \mwci/ have a maximum leakage score.

Additionally, \mwci/ discovers input-dependent behavior in the AES-GCM encryption of node-forge. Manual inspection shows that these leakages in the \texttt{tableMultiply} function in the file \texttt{cipherModes.js} occur during the computation of the GHASH which is used for the final computation of the authentication tag. The \texttt{tableMultiply} function uses a table precomputed from the hash key and multiplies by accessing this table with an index which is an intermediate value computed from the current ciphertext block and the previous hash value. Learning this intermediate value potentially allows to gain information about the GHASH key, compromising the authentication property. The implementation in \texttt{node-forge} uses 4-bit tables. Whether this implementation and leakage is exploitable, is left to future work. We recommend not having any secret-dependent non-constant-time code.

\subsubsection{Elliptic curve implementations}
\texttt{node-forge} and \texttt{tweetnacl} feature custom constant-time big number arithmetic that is specifically designed for the supported curves. 
The \texttt{elliptic} library, however, relies entirely on arithmetic from the general-purpose \texttt{bn.js}~\cite{lib-bn.js} library, which features a lot of input-dependent control flow and memory accesses. 
Thus, we see very high leakage over all supported primitives. The leakages detected in the big number and elliptic code itself are mostly assigned scores between 80 and 100.

In addition, for computing the signature, \texttt{elliptic}'s ECDSA implementation uses the \texttt{hash.js}~\cite{lib-hash.js} library, which offers pure-\javascript/ implementations for SHA-1 and SHA-2. 
For ECDSA and EdDSA signatures with the curves \texttt{p384} and \texttt{ed25519}, respectively, the leakage report points to a significant amount of leakage in \texttt{lib/hash/sha/512.js} for a variety of call stacks. 
Here, the implementation works around a limitation of \javascript/, which represents all numbers in IEEE-754 double precision floating point, and temporarily converts them to 32-bit signed integers for bitwise arithmetic. If the most-significant bit ends up being 1, \javascript/ sign-extends it such that the result is negative. The implementation checks for this in an \texttt{if} statement and adds \texttt{0x100000000} to get a positive number.
This leakage may pose a security issue, as ECDSA and EdDSA use the hash function for generating a nonce from the private key. \mwci/ assigns leakage scores between 60 and 70 for most of the leakages in \texttt{lib/hash/sha/512.js}. Future work could investigate whether the leakage of the most-significant bit can be used to learn parts of the private key.
The libraries \texttt{elliptic}, \texttt{bn.js} and \texttt{hash.js} are from the same author.

\subsubsection{Base64 encoding}
We also found leakages in some of the various Base64 implementations. 
All of them were caused by the use of lookup tables, where 6-bit chunks are mapped to ASCII characters and vice versa.
The only known attack against Base64 encoding relies on a precise controlled channel that is not available for common \javascript/ deployments~\cite{DBLP:conf/ccs/SieckBW021}. However, depending on the memory layout of the respective lookup tables, partial information may be accessible via a cache attack.
\texttt{js-base64} does also feature a vulnerable Base64 implementation; however, it first checks whether the \texttt{Buffer} class with native Base64 support is present, which is the case for our \nodejs/ build.

\subsection{Number of Test Cases}
\label{sec:discussion:testcases}
As mentioned in the performance analysis, computation time and, to a lesser degree, memory consumption, scale with the number of test cases. A higher number of test cases increases the chance of triggering uncommon code paths and thus finding more leakages. In the following, we analyze this trade-off and point out approaches for striking a good balance between accuracy and performance.

In our performance analysis, we ran 16 test cases for each library. This number is within the same order of magnitude as the one used for the evaluation in~\cite{weiser2018data}, where the authors recommend running 10 test cases. To check whether the small number of test cases had impact on the number of detected leakages, we repeated our analysis with 48 additional test cases (64 total) for each target and compared the results with those of the first analysis.

\subsubsection{Performance}

Increasing the number of test cases does not affect every pipeline step in the same way. 
Doubling the number of test cases roughly doubles the CPU time needed for trace generation, but that does not apply to the analysis step: There, the first test case takes much longer than subsequent ones, as it needs to build the tree from scratch, which involves spending a lot of time in the memory allocator. 
Later non-diverging test cases only need to iterate the existing tree, which takes considerably less resources.
We observed that the duration increased by factor 3 to 3.5, although we ran 4 times as many test cases.

\subsubsection{Leakages}
Except for targets in the libraries \texttt{elliptic} and \texttt{node-forge}, \mwci/ found the same amount of leakages with 64 test cases as with 16. For \texttt{elliptic}, all targets show a small single digit increase in the number of overall and unique leakages. For all new leakages, we determined that these were initially missed due to a saturation effect (see Section \ref{sec:limitations}) and not by lack of coverage, and would have been found by re-running the analysis after fixing the preceding leakages.

For \texttt{node-forge}'s RSA implementation, the difference is a bit larger. While \mwci/ finds 223 overall and 111 unique leakages with 16 test cases, it was able to discover 255 overall and 125 unique leakages with 64 test cases. Manual investigation shows again that most leakages were missed due to a saturation effect. However, a small number was missed due to insufficient coverage of the initial 16 test cases.

\subsubsection{Recommendations}
We recommend the developer to choose an overall duration that is acceptable during ongoing development and determine an according test case number.
In addition, the coverage of the generated test cases could be checked with a separate tool to ensure that all relevant code gets executed. Finally, the developer could add another larger collection of test cases that runs as a final check 
before releasing the next version, where a longer analysis time is acceptable.

\subsection{Comparison with Microwalk's original Analysis Module}
\label{sec:discussion:old-microwalk-analysis}
\mw/ originally features two analysis modules that implement the \textit{memory access trace (MAT)}~analysis method for finding leakages. 
The method was first presented in~\cite{wichelmann2018microwalk}. %
For each memory accessing instruction, the modules generate a hash over all accessed offsets. 
By comparing the hashes between traces, the amount of leakage for each memory accessing instruction is computed. Due to the focus on memory accesses, control flow leakages are only discovered indirectly or may even be missed entirely.

The first module, that was originally published with~\cite{wichelmann2018microwalk}, generates only one leakage report per instruction.
The later added second module (referred to by us as \textit{CMAT module}) is an extension of the first module that additionally distinguishes between call stacks to achieve a higher accuracy. To compare the existing analysis method with our new approach, we ran a selection of the targets with the CMAT module, using the same 16 test cases as for the initial analysis. The results are shown in Table~\ref{tab:js-mw-legacy-analysis}.

\begin{table}[t]
    \caption{Results of the analysis step of selected targets with the original \mw/ CMAT module, and its resource usage. Time and memory consumption of the trace generation and preprocessing steps are identical to those shown in Table~\ref{tab:js-mw-libs}.}
    \label{tab:js-mw-legacy-analysis}
    \centering
    \begin{tabular}{ l r r r r r }
        Target & CPU & Duration & RAM & \# Lkgs. & \# Unique \\ \hline \hline
        
        \multicolumn{2}{l}{\textbf{aes-js}} & & & & \\
            AES-ECB & $<$ 1 sec & 8 sec & 168 MB & 16 & 16 \\
        \hline
        
        \multicolumn{2}{l}{\textbf{elliptic}} & & & & \\
            p192 & 4 sec & 253 sec & 289 MB & 4,003 & 811 \\
        \hline
                
        \multicolumn{2}{l}{\textbf{tweetnacl}} & & & & \\
            ed25519 & 5 sec & 126 sec & 286 MB & 0 & 0 \\
        \hline
    \end{tabular}
\end{table}

Since the CMAT module only stores a single mapping of call stacks and instructions to hashes, it generally takes less resources than our new tree-based approach, both in computation time and memory consumption. However, the preceding trace generation and preprocessing, which take most of the time, are identical, so the actual difference in overall duration is limited.

For \texttt{aes-js}' AES-ECB implementation, the CMAT module reports a number of secret-dependent table accesses with full leakage, which are identical to the leakages reported by our new analysis module.
This is the kind of leakage that the MAT analysis was designed for: 
Through hashing the sequence of memory addresses that a given instruction accesses, secret-dependent variations are discovered. 
Our new analysis detects these leakages through the address lists stored in the individual memory access trace entries, which ultimately yields the same result, but takes more memory.

The result from the CMAT module for \texttt{elliptic}'s p192 is very imprecise and contains many false positives: It reports 811 leaking lines in total, which includes lines like ``\lstinline[basicstyle=\small\ttfamily,breaklines=true]|this.pendingTotal = 0;|''. As a fixed offset is accessed, this line is a clear false positive. The leakage in question was in fact caused by a control flow variation higher up in the call chain, leading to a varying number of executions of the given instruction, which in turn produced a different memory access offset hash. The other false positives follow a similar pattern.
Our new tree-based approach handles control flow and memory access leakages separately, which reduces false positives and allows accurately attributing a leakage to a specific code line.

\subsection{Limitations of the Analysis Algorithm}
\label{sec:limitations}
As other dynamic analysis approaches, \mwci/ needs a good coverage of the program in order to give an accurate leakage detection result. If a particular path is never executed, it does not appear in the traces and thus never reaches the analysis modules. However, for cryptographic code, randomly generated test cases tend to work very well~\cite{wichelmann2018microwalk,weiser2018data}. For other targets, it may be worth exploring other methods for generating coverage, e.g., fuzzing.

Finally, in our analysis algorithm, some leakages may be obscured by other leakages at a higher tree level. If leakages on higher levels cause splits that result in a unique sub tree for each trace, the lower leakages can not cause any more divergences and thus are overlooked. 
This ``saturation'' is an inherent property of the analysis approach, and the price payed for having a linear-time algorithm. 
We do not believe that this impacts practical usage: After having a library reach a certain state of ``constant-time-ness'', we only expect few new leakages being reported, as certain functions are touched. 
And even if a leakage is not reported in a first pass, it will show up after committing the fixes for the previously reported leakages. It is unlikely that a number of unfixed low-severity leakages obscure a subsequent severe leakage. This would imply a fully split up tree, which, in itself, signals a high-severity leakage.

Other work tries to find all trace leakages in a single pass,
but uses significantly more resources with every CI run and thus is not suitable for integration into an everyday-development workflow.

\section{Related Work}
\label{sec:related-work}

\bheading{Constant-time program analysis} has a long tradition as there are different classes of vulnerabilities that can be found through various analysis techniques~\cite{DBLP:journals/csur/LouZJZ21}. 
Some tools for checking constant-time behavior depend on the availability of source code.
Irazoqui et al.~\cite{DBLP:journals/corr/abs-1709-01552} introduce secret-dependent cache trace analysis, \emph{ct-fuzz}~\cite{DBLP:conf/icst/HeEC20} specializes fuzzing for timing leakages, \emph{ct-verif}~\cite{DBLP:conf/uss/AlmeidaBBDE16} describes constant-time through safety properties and \emph{CaSym}~\cite{DBLP:conf/sp/BrotzmanLZTK19} uses symbolic execution to model the execution behavior of a program.
\mwci/ does not require access to the source code for compiled languages. 

Unlike \mwci/ which uses dynamic program analysis and compares real execution traces, \textbf{static binary analysis} tries to simulate the execution of every possible program path.
\emph{BINSEC/REL}~\cite{DBLP:conf/sp/DanielBR20} uses relational symbolic execution of two execution traces %
to efficiently analyze binary code, however is limited by the high performance impact of static analysis.
\emph{CacheS}~\cite{DBLP:conf/uss/0011BL0ZW19}, based on \emph{CacheD}~\cite{DBLP:conf/uss/WangWLZW17}, combines taint tracking and symbolic execution to find cache line granular leakage and secret-dependent branches.
Moreover, \emph{CacheAudit}~\cite{DBLP:conf/uss/DoychevFKMR13} tracks relational information about memory blocks to compute upper bounds for leakages. 
In contrast with these works, \mwci/ finds any leakage with byte granularity.
\DATA/~\cite{weiser2018data} and its (EC)DSA-specific extension~\cite{DBLP:conf/uss/WeiserSBS20} find microarchitectural and timing side-channels in binaries via \textbf{dynamic binary analysis}.
The trace alignment approach of \DATA/ is based on computing pairwise differences between traces, leading to a computation time that is quadratic both in the number of traces and in the trace length. While it yields more leakage candidates after a single pass, it needs more computational resources and thus is not a suited for use in a CI environment.
\emph{Abacus}~\cite{DBLP:conf/icse/BaoWLLW21}
identifies secret-dependent memory access instructions using symbolic execution.
Then, the authors use Monte Carlo sampling to estimate the amount of leaked information.
A shortcoming of the approach is that \emph{Abacus} only uses one trace and therefore suffers from low coverage.
\emph{dudect}~\cite{DBLP:conf/date/ReparazBV17} 
measures timing behavior in a statistical way without any model of the underlying hardware, which is fast, but also yields imprecise results.
\emph{ctgrind}~\cite{langley2010ctgrind} and \emph{TIMECOP}~\cite{neikes2020timecop} search the code for secret-dependent jump or memory accesses like table-lookups and variable-time CPU instructions, but are rather manual.
\bheading{Analysis of \javascript/ code} recently received more focus in the research community as it is widely used in browsers including many security-critical workloads.
Basic properties of \javascript/ regarding security of code have been widely analyzed~\cite{DBLP:conf/sas/JensenMT09,DBLP:conf/sp/TalyEMMN11,DBLP:conf/sigsoft/SenKBG13,DBLP:conf/sigsoft/KashyapDKWGSWH14}.
Just as in other programming languages, various attacks on secret-dependent behavior have been conducted~\cite{DBLP:conf/fc/SchwarzMGM17,DBLP:conf/uss/ShustermanAOGOY21}.
A common prerequisite for exploiting timing-dependent properties of code is having precise timers~\cite{DBLP:conf/eurosp/RokickiML21},
though this can be bypassed~\cite{DBLP:conf/uss/ShustermanAOGOY21}.
Apart from countermeasures like disabling timers or blocking certain functionality~\cite{DBLP:conf/ndss/0001LG18}, little work has gone into finding non-constant-time \javascript/ code.

\section{Conclusion}
\label{sec:conclusion}

With \mwci/ we have shown how one can design a side-channel analysis framework that is suitable for integration into a day-to-day development workflow. We have presented a new trace processing algorithm that merges the recorded traces into a call tree, allowing us to precisely localize and quantify leakages in a short time frame.
Moreover, by ``dockerizing'' the analysis, we have provided the means for easy and fast usage without the necessity of understanding the details of the framework.

With the design and implementation of a tracer for \javascript/ and the integration with \mwci/, we have built the first comprehensive constant-time verifier for \javascript/ code and demonstrated how analysis techniques originally developed for binary analysis can be used for interpreted or just-in-time compiled
languages. \mwci/ is constructed in a modular fashion and allows to add tracing backends for other languages with limited effort.

Overall, \mwci/ carries the potential to increase the side-channel security for many popular libraries written in potentially any programming language, and raises awareness for the risks of non-constant-time code in new communities.

\begin{acks}
The authors thank Julia Tönnies for her help in evaluating the suitability of the leakage metrics, and the anonymous reviewers for their helpful comments and suggestions. This work has been supported by Deutsche Forschungsgemeinschaft (DFG) under grants 427774779 and 439797619, and by Bundesministerium für Bildung und Forschung (BMBF) through projects ENCOPIA and PeT-HMR.
\end{acks}

\newpage

\bibliographystyle{ACM-Reference-Format}
\balance
\bibliography{references}

\newpage

\begin{appendix}

\section{Call tree dump}
\label{sec:app:call-tree-dump}

\begin{figure}[H]
\begin{mdframed}[style=listingframe]
\begin{lstlisting}[basicstyle=\linespread{1.1}\scriptsize\ttfamily,language=]
(*@\indentrule@*)(*@\color{halfgreen}{\textbf{@root}}@*)
(*@\indentrule@*)  (*@\color{halfred}{Trace entries:}@*)
(*@\indentrule@*)    (*@\indentrule@*)(*@\color{blue}{\textbf{\#call}}@*) main+1 -> func+0
(*@\indentrule@*)    (*@\indentrule@*)  (*@\color{halfred}{Trace entries:}@*)
(*@\indentrule@*)    (*@\indentrule@*)    (*@\indentrule@*)(*@\color{blue}{\textbf{\#call}}@*) func+1 -> lookup+0
(*@\indentrule@*)    (*@\indentrule@*)    (*@\indentrule@*)  (*@\color{halfred}{Trace entries:}@*)
(*@\indentrule@*)    (*@\indentrule@*)    (*@\indentrule@*)    (*@\indentrule@*)(*@\color{blue}{\textbf{\#memory-read}}@*) at lookup+1
(*@\indentrule@*)    (*@\indentrule@*)    (*@\indentrule@*)    (*@\indentrule@*)  table[1]: (*@\color{red}{0}@*)
(*@\indentrule@*)    (*@\indentrule@*)    (*@\indentrule@*)    (*@\indentrule@*)  table[2]: (*@\color{red}{1}@*)
(*@\indentrule@*)    (*@\indentrule@*)    (*@\indentrule@*)    (*@\indentrule@*)  table[3]: (*@\color{red}{2}@*)
(*@\indentrule@*)    (*@\indentrule@*)    (*@\indentrule@*)    (*@\indentrule@*)(*@\color{blue}{\textbf{\#return}}@*) lookup+1 -> func+1
(*@\indentrule@*)    (*@\indentrule@*)    (*@\indentrule@*)(*@\color{blue}{\textbf{\#jump}}@*) func+4 -> <?> (not taken)
(*@\indentrule@*)    (*@\indentrule@*)    (*@\indentrule@*)(*@\color{blue}{\textbf{\#call}}@*) func+5 -> lookup+0
(*@\indentrule@*)    (*@\indentrule@*)    (*@\indentrule@*)  (*@\color{halfred}{Trace entries:}@*)
(*@\indentrule@*)    (*@\indentrule@*)    (*@\indentrule@*)    (*@\indentrule@*)(*@\color{blue}{\textbf{\#memory-read}}@*) at lookup+1
(*@\indentrule@*)    (*@\indentrule@*)    (*@\indentrule@*)    (*@\indentrule@*)  table[1]: (*@\color{red}{0, 1, 2}@*)
(*@\indentrule@*)    (*@\indentrule@*)    (*@\indentrule@*)    (*@\indentrule@*)(*@\color{blue}{\textbf{\#return}}@*) lookup+1 -> func+5
(*@\indentrule@*)    (*@\indentrule@*)    (*@\indentrule@*)(*@\color{blue}{\textbf{\#jump}}@*) func+6 -> func+4
(*@\indentrule@*)    (*@\indentrule@*)  (*@\color{halfred}{Splits:}@*)
(*@\indentrule@*)    (*@\indentrule@*)    (*@\indentrule@*)(*@\color{halfgreen}{\textbf{@split}}@*): (*@\color{red}{0}@*)
(*@\indentrule@*)    (*@\indentrule@*)    (*@\indentrule@*)  (*@\color{halfred}{Trace entries:}@*)
(*@\indentrule@*)    (*@\indentrule@*)    (*@\indentrule@*)    (*@\indentrule@*)(*@\color{blue}{\textbf{\#jump}}@*) func+4 -> func+7
(*@\indentrule@*)    (*@\indentrule@*)    (*@\indentrule@*)    (*@\indentrule@*)(*@\color{blue}{\textbf{\#return}}@*) func+7 -> main+1
(*@\indentrule@*)    (*@\indentrule@*)    (*@\indentrule@*)(*@\color{halfgreen}{\textbf{@split}}@*): (*@\color{red}{1, 2}@*)
(*@\indentrule@*)    (*@\indentrule@*)    (*@\indentrule@*)  (*@\color{halfred}{Trace entries:}@*)
(*@\indentrule@*)    (*@\indentrule@*)    (*@\indentrule@*)    (*@\indentrule@*)(*@\color{blue}{\textbf{\#jump}}@*) func+4 -> <?> (not taken)
(*@\indentrule@*)    (*@\indentrule@*)    (*@\indentrule@*)    (*@\indentrule@*)(*@\color{blue}{\textbf{\#call}}@*) func+5 -> lookup+0
(*@\indentrule@*)    (*@\indentrule@*)    (*@\indentrule@*)    (*@\indentrule@*)  (*@\color{halfred}{Trace entries:}@*)
(*@\indentrule@*)    (*@\indentrule@*)    (*@\indentrule@*)    (*@\indentrule@*)    (*@\indentrule@*)(*@\color{blue}{\textbf{\#memory-read}}@*) at lookup+1
(*@\indentrule@*)    (*@\indentrule@*)    (*@\indentrule@*)    (*@\indentrule@*)    (*@\indentrule@*)  table[1]: (*@\color{red}{1, 2}@*)
(*@\indentrule@*)    (*@\indentrule@*)    (*@\indentrule@*)    (*@\indentrule@*)    (*@\indentrule@*)(*@\color{blue}{\textbf{\#return}}@*) lookup+1 -> func+5
(*@\indentrule@*)    (*@\indentrule@*)    (*@\indentrule@*)    (*@\indentrule@*)(*@\color{blue}{\textbf{\#jump}}@*) func+6 -> func+4
(*@\indentrule@*)    (*@\indentrule@*)    (*@\indentrule@*)  (*@\color{halfred}{Splits:}@*)
(*@\indentrule@*)    (*@\indentrule@*)    (*@\indentrule@*)    (*@\indentrule@*)(*@\color{halfgreen}{\textbf{@split}}@*): (*@\color{red}{1}@*)
(*@\indentrule@*)    (*@\indentrule@*)    (*@\indentrule@*)    (*@\indentrule@*)  (*@\color{halfred}{Trace entries:}@*)
(*@\indentrule@*)    (*@\indentrule@*)    (*@\indentrule@*)    (*@\indentrule@*)    (*@\indentrule@*)(*@\color{blue}{\textbf{\#jump}}@*) func+4 -> func+7
(*@\indentrule@*)    (*@\indentrule@*)    (*@\indentrule@*)    (*@\indentrule@*)    (*@\indentrule@*)(*@\color{blue}{\textbf{\#return}}@*) func+7 -> main+1
(*@\indentrule@*)    (*@\indentrule@*)    (*@\indentrule@*)    (*@\indentrule@*)(*@\color{halfgreen}{\textbf{@split}}@*): (*@\color{red}{2}@*)
(*@\indentrule@*)    (*@\indentrule@*)    (*@\indentrule@*)    (*@\indentrule@*)  (*@\color{halfred}{Trace entries:}@*)
(*@\indentrule@*)    (*@\indentrule@*)    (*@\indentrule@*)    (*@\indentrule@*)    (*@\indentrule@*)(*@\color{blue}{\textbf{\#jump}}@*) func+4 -> <?> (not taken)
(*@\indentrule@*)    (*@\indentrule@*)    (*@\indentrule@*)    (*@\indentrule@*)    (*@\indentrule@*)(*@\color{blue}{\textbf{\#call}}@*) func+5 -> lookup+0
(*@\indentrule@*)    (*@\indentrule@*)    (*@\indentrule@*)    (*@\indentrule@*)    (*@\indentrule@*)  (*@\color{halfred}{Trace entries:}@*)
(*@\indentrule@*)    (*@\indentrule@*)    (*@\indentrule@*)    (*@\indentrule@*)    (*@\indentrule@*)    (*@\indentrule@*)(*@\color{blue}{\textbf{\#memory-read}}@*) at lookup+1
(*@\indentrule@*)    (*@\indentrule@*)    (*@\indentrule@*)    (*@\indentrule@*)    (*@\indentrule@*)    (*@\indentrule@*)  table[1]: (*@\color{red}{2}@*)
(*@\indentrule@*)    (*@\indentrule@*)    (*@\indentrule@*)    (*@\indentrule@*)    (*@\indentrule@*)    (*@\indentrule@*)(*@\color{blue}{\textbf{\#return}}@*) lookup+1 -> func+5
(*@\indentrule@*)    (*@\indentrule@*)    (*@\indentrule@*)    (*@\indentrule@*)    (*@\indentrule@*)(*@\color{blue}{\textbf{\#jump}}@*) func+6 -> func+4
(*@\indentrule@*)    (*@\indentrule@*)    (*@\indentrule@*)    (*@\indentrule@*)    (*@\indentrule@*)(*@\color{blue}{\textbf{\#jump}}@*) func+4 -> func+7
(*@\indentrule@*)    (*@\indentrule@*)    (*@\indentrule@*)    (*@\indentrule@*)    (*@\indentrule@*)(*@\color{blue}{\textbf{\#return}}@*) func+7 -> main+1
\end{lstlisting}
\end{mdframed}
\caption{Call tree dump for the example in Figure~\ref{fig:varying-loop-iterations} and three different values of \texttt{secret}: \texttt{1} (trace ID \texttt{0}), \texttt{2} (trace ID \texttt{1}) and \texttt{3} (trace ID 
\texttt{2}). The dump is generated by running a depth-first search on the tree and printing the individual nodes with appropriate indentation. Trace entry types are highlighted with blue color, trace IDs with red color.}
\label{fig:call-tree-dump}
\end{figure}

\pagebreak

\end{appendix}

\end{document}